\documentclass[letterpaper,11pt]{article}

\usepackage[y4project]{edmaths} 
\usepackage[utf8]{inputenc}
\usepackage{mathpazo}
\usepackage{setspace} 
\usepackage{tabularx} 
\usepackage{tocbibind}
\usepackage{amsmath}  
\usepackage{graphicx} 
\usepackage[margin=1in,letterpaper]{geometry} 
\usepackage{url}
\usepackage{hyperref}
\usepackage{xcolor}
\hypersetup{
    colorlinks,
    linkcolor={blue!50!black},
    citecolor={blue!50!black},
    urlcolor={blue!80!black}
}
\usepackage[natbib=true,backend=biber,sorting=nyt,style=apa]{biblatex}

\begin{document}
    \begin{center}

        \vspace*{1cm}
            
        \Huge
        \textbf{Quantum Annealing for the Set Splitting Problem}
        
        \vspace{0.5cm}

        \vspace{1.5cm}
        \large
         \textbf{Sean Borneman \\Bloomington High School South, IN}
            
        \vfill

    \end{center}

\pagenumbering{roman}

\newpage
\tableofcontents
\newpage
\pagenumbering{arabic}

\newpage

\section{Introduction}
This paper presents a new algorithm for Set Splitting and Max Set Splitting problem (also known as the Hypergraph Coloring Problem). The Set Splitting problem states: Given a family F of subsets of a finite set S, decide whether there exists a partition of S into two subsets S1, S2 such that all elements of F are split by this partition, i.e., none of the elements of F is completely in S1 or S2. The Optimization version, The Max Set Splitting Problem, is the optimization variant of the problem which asks for the Split which maximizes the number of subsets in $F$ that are split by the partition of $S_1$ and $S_2$. There are a few special cases of this problem. If each set in $F$ is restricted to a cardinalityy of $k$, i.e each subset in $F$ has $k$ elements, it is referred to as the k-Set-Splitting Problem, a special case of this - where $k=2$ - is know as the Max-Cut problem (\cite{guruswami2004inapproximability}). This paper attempts to leverage quantum computing which has the potential to dramatically reduce the time and number of operations, required to solve the Set Splitting problem. 

Conventionally, the k-set-splitting problem has been simplified with kernalization algorithms for a time complexity of $O(1.9630^{n}+N)$ where $n$ is the number of elements and $N$ is the size of the instance (\cite{lokshtanov2009even}).  Quantum annealing has the potential to out pace such conventional methods, while also providing exact solutions, with it's low time complexity of  $O(e^{\sqrt{n}})$  where $n$ is the problem size (number of qubits) (\cite{mukherjee2015multivariable}).

Quantum annealing, an adiabatic method of quantum computing, uses quantum fluctuations to find the global minimum of a given function. Quantum annealing is used to solve optimization problems, such as RNA folding, Portfolio Optimization, as well as the Hamiltonian Cycle problem and Shortest Vector Problem (\cite{mcgeoch2019practical}). The current largest Quantum Annealer hardware, the D-Wave Advantage, can support up to $5760$ qubits. It has already been proven that quantum annealing can solve the Hamiltonian cycle problem (\cite{mahasinghe2019solving}), and the traveling salesman problem (\cite{matai2010traveling}) with potential to outpace classical solutions, it does however, remain to be seen if a quantum solution to the Set Splitting Problem exists.

 \section{Background}
The Set Splitting Problem is an NP-Complete problem meaning that it cannot be solved by a classical computer in polynomial time, however a given answer can be verified in polynomial time (i.e. the time to solve scales polynomially with the number of elements). This work presents a novel quantum algorithm for solving the Set Splitting and Max Set Splitting Problem designed for the adiabatic framework of quantum computation. The adiabatic framework is used because it allows for more qubits to be used for calculations than can currently be used on the best available quantum circuit computers which currently have a maximum of $1000$ qubits, meaning that although they allow for more control, since individual logic gates in the form of matrix operations can be placed on each qubit, the size of the problems they can currently solve are much smaller than on a Quantum Annealer.

Regarding applications, the Set Splitting problem has a variety of commercial and research applications; most notably in the analysis of Micro-array data, a data set of DNA and gene expression requiring computer analysis (\cite{chen2009improved}). When this data is viewed as a matrix of samples, then DNA set splitting can be used to form a rough hypothesis for the genes responsible for the expression of various phenotypes, such as for the emergence of cancer tumors (\cite{dehne2003fpt}). The Set Splitting problem can also have benefits to cybersecurity and network organization when used as a graph coloring algorithm. In the domain of  cybersecurity, servers, endpoints, and network nodes can be modelled as graph vertices, with connectivity edges representing attack vectors. Solving a set-splitting problem for such a representation can allow one to partition security controls to isolate risks. Networks of multiple high permission computers leave themselves open to Stuxnet type attacks where a single computer has elevated permissions on a large network. By splitting this network such that susceptible nodes are in separate fire-walled subnets we can improve the security of a organizations network.

Quantum annealing uses quantum fluctuation to solve optimization problems. As opposed to simulated annealing (a probabilistic algorithm runnable on a conventional computer) which relies on modeling thermal fluctuations to find the global minimum, Quantum annealing takes advantage of quantum fluctuations such as quantum tunneling to transition between local minima (\cite{ray1989sherrington}). Quantum annealing can be more efficient than simulated annealing thermal fluctuations, because conventionally, high energy barriers between local minima are extremely hard to overcome and simulated annealing therefore can only address one possible solution at a time, which leads to an increase in time as the problem space increase. However in quantum annealing the entire problem space is seen at once, leading to much faster ground state (global optimization) calculations (\cite{das2005quantum}).

Quantum Annealing relies on finding the energetic ground state of a Hamiltonian where the ground state of that Hamiltonian corresponds to the solution to the computation. Quantum Annealers, the hardware responsible for this process, encode the coefficients of some Ising Hamiltonian or quadratic unconstrained binary optimization (QUBO) onto the qubits of a quantum computer which then returns the solution vector $z$ that minimizes the energy of the Hamiltonian (see equation \ref{Ham}):

\begin{equation} \label{Ham}
H(z) =\sum_{i=1}^{L}Q_{i,i}z_{i}+\sum_{i<j}^{L} Q_{i,j} z_{i} z_{j}
\end{equation}

where $L$ is the number of qubits and $z$ is the solution vector where  $z_i \in {-1, 1}$ for Ising Hamiltonian's, and $z_i \in {0,1}$ for QUBO formulations, the latter of which will be used for the remainder of this paper (\cite{mcgeoch2020theory}).

The largest downsides of the current generation of quantum annealers are mapping, and solution reliability. The mapping issue exists because the modern topologies of quantum annealers do not have perfect connectivity between every possible logical qubit. In the current generation of D-wave Advantage QPU's the Pegasus architecture allows for qubits to have a nominal length of 12 and a degree of 15 meaning each qubit is connected to 12 orthogonal qubits through internal couplers, and 15 qubits through internal and external couplers.(\cite{boothby2020next}). In practice, this means that despite some example problem only theoretically requiring $75$ logical qubits (ones expressed in the problem Hamiltonian and which directly determine the solution), it may actually require $340$ to be encoded onto the QPU, because of the non-perfect inter-connectivity. Separately, the issue of solution reliability is inherent to adiabatic solvers, but it does mean that on any given calculation, the final output may not correspond to the ground state of the problem Hamiltonian (\cite{boixo2014evidence}).

\section{Quantum Annealing formulation for Set Splitting problem} 
I present a method of encoding the Set Splitting, and Max Set Split problem as a Hamiltonian H, such that the energetic ground state corresponds to the $S_1$ and $S_2$ solutions. I also present the results of this encoding method run on a D-Wave Quantum Annealer. The approach encodes each subset inS $F$ as a penalty function to be optimized. The D-Wave Advantage provides native support for QUBO implementation as well as data on the required qubits and on quantum compute access time.

\subsection{QUBO Formulation}

The Set Splitting problem is defined as: Given a family F of subsets of a finite set S, decide whether there exists a partition of S into two subsets S1, S2 such that all elements of F are split by this partition, i.e., none of the elements of F is completely in S1 or S2.

The QUBO formulation begins by constructing 2 matrices, $F$ which encodes sets of elements as described above and $X$ which encodes the two subsets $S1$ and $S2$. 

$X$ is encoded such that for any possible matrix $X$ $$X_{i} = 1 \text{ if and only if element } i \text{ is contained within subset } S_1$$ This also means that if $X_i=0$, then element i is contained within subset S2. This construction of $X$ contains a binary number which represents each element, where its value denotes in which subset ($S_1$ or $S_2$) the respective element is contained. For example given a finite set $S$ and family of subsets $F $. 
$$ S =
  \left[ {\begin{array}{ccccc}
    A & B & C & D & E
  \end{array} } \right]
$$
$$
  F =
  \left[ {\begin{array}{ccccc}
    A & B\\
    B & D \\
    A & C& E\\
  \end{array} } \right]
$$

A clear solution is $S_1 = (B, C, E)$, $S_2 = (A, D)$ and it could be encoded in the matrix $X=[0 1 1 0 1]$ or  $X=[1 0 0 1 0]$. 

Given this formulation there is a single condition for the solution to be valid:

$$2. \text{ Neither } S_1 \text{ nor } S_2 \text{ completely contains any subset within } F$$

For context given the previously established $S$ and $F$  a split such as $S_1 = (a_2, a_3, a_4)$, $S_2 = (a_1, a_5)$ would not be valid since the subset of $F $ $[a_2,a_4]$ would be completely contained within $S_1$. The penalty conditions for this 'rule' is then:
\begin{equation} \label{H1}
H_1(X) = \sum_{j=0}^{n-1} \frac{1}{m-1}\sum_{i_0=0}^{m-1} \sum_{i_1=i_0+1}^{m-1} X_{F(j,i_0)} X_{F(j,i_1)}
\end{equation}

\begin{equation} \label{H2}
H_2(X) = \sum_{j=0}^{n-1} \frac{1}{m-1}\sum_{i_0=0}^{m-1} \sum_{i_1=i_0+1}^{m-1} (1-X_{F(j,i_0)})(1-X_{F(j,i_1)})
\end{equation}

Where $n$ denotes the number of subsets in $F$, $m$ denotes the length of the $j$th subset in $F$, and $F(i,j)$ denotes the $i$th element in the $j$th subset of F. The full Hamiltonian can then be expressed as a sum of the two penalty conditions:
$$H(x) = H_{1}(x) + H_{2}(x)$$

The penalty function $H_1$  multiplies every pair of elements (denoted by an index of X) in a subset together, the result being that the minimum occurs when every element in a particular subset of $F$ is not present in subset $S_1$ i.e. not all $x$ indices denoted by the elements in that subset are $1$. $H_2(X)$ does the same but ensures that the elements in each pair are not all $0$. It works the same way $H_1(x)$ does but substitutes each variable $X$ for $1-X$, this inversion allows us to check for when every element in a particular subset $F$ is not present in subset $S_2$. $H_2(x)$ is necessary because $S_1$ and $S_2$ are both denoted by the elements in $X$ where an element being denoted by the value of $0$ denotes it's representation in one set while a value of $1$ denotes its presents in the other. Therefore, it is necessary to check that both the subset denoted by $0$ and the subset denoted by $1$ does not fully contain a subset of $F$. 

The reason that we can encode a subset of more than $3$ elements without increasing the depth of the resulting Binary Optimization Function is as follows. Consider an example subset of $F$: $(a_1,a_2,a_3)$; it doesn't matter whether we encode the minimum as $x_1x_2 + x_1x_3+x_2x_3$ or $x_1x_2x_3$ because the values of $x$ at which the function is at a minimum is the same despite the overall function being different. Crucially the difference between two of the three variables being $1$ and all three of the variables being $1$ is the same, once normalized by a factor of $\frac{1}{2}$. Because of this property we can substitute the $x_1x_2x_3$ term for $x_1x_2+x_1x_3+x_2x_3$ in order to prevent increasing the chain length while still tracking the $3$ representative qubits.

The issue with the current approach  is that if the algorithm can reduce the number of elements in a set $F$ that end up in a set $S$ this optimizes the function, potentially at the expense of the number of $F$ subsets that get split. For example consider some input set $F$:
$$
  F =
  \left[ {\begin{array}{ccccc}
    a_1 & a_2 & a_3 & a_4 & a_5\\
   a_2 & a_3 & a_4 & a_5 & a_6\\
    a_1 & a_5 \\
    a_2 & a_5\\
    a_3 & a_5 \\
    a_4 & a_5\\
    a_6 & a_5 \\
  \end{array} } \right]
$$
The optimization of splitting the first $F$ subset into a set of $3$ and a set of $2$, saving $3/4$ when compared to a $4$, $1$ split. This optimization of larger k sets can override the number smaller k sets that are truly split, if enough of these larger set size optimizations are made.  In the above $F$, the correct split of the $k=2$, subsets may be sacrificed in exchange for a $3$,$2$ split in both of the two $k=5$ subsets. This is an issue with the overall algorithm and it is a problem with inputs $F$ that have a large number of solutions with a lot of subsets with a high cardinality. Beneficially, it is less likely to occur for inputs with a low number of solutions as any decreased output gain from optimizing the split of a set is offset by the breaking of a condition. This could be solved by increasing the logical chain length ($2$ in a QUBO format).

 \subsection{Variations}
The $3$ common variations, Max-Set-Split, Weighted Set Split and K-Set-Split - special cases of the Set-Splitting Problem - are trivial once we lay the ground work for the generalized Problem. The K-Set-Split problem is already solvable with the current formulation, the difference being that $m$ will always be some constant $k$. The Max-Set-Split variation is also already solved, because the penalty conditions decrease by $1$ for each condition left unbroken. The Weighted Set Split problem adds a condition: each subset in the collection F is associated with a weight that is a real number, and the objective is to construct a partition of the ground set that maximizes the sum of the weights of the split subsets.

 We can solve this Weighted Set Split variation by adding a multiplier to each set:

 \begin{equation} \label{H3}
 H_1(X) = \sum_{j=0}^{n-1} \frac{W_{j}}{m-1}\sum_{i_0=0}^{m-1} \sum_{i_1=i_0+1}^{m-1} X_{F(j,i_0)} X_{F(j,i_1)}
\end{equation}

\begin{equation} \label{H4}
H_2(X) = \sum_{j=0}^{n-1} \frac{W_{j}}{m-1}\sum_{i_0=0}^{m-1} \sum_{i_1=i_0+1}^{m-1} (1-X_{F(j,i_0)})(1-X_{F(j,i_1)})
\end{equation}

where $W_{j}$ is the weight of the $j$th subset. Unfortunately this does not reliably work for $k > 3$ because the previously mentioned split optimization issue changes such that a single split optimization on a set with a large weight could override a condition. Since this issue occurs for $k > 3$ this weighted solution only applies for $k \le 3$
\subsection{Example}
In this section I will show how this can be applied to an example Set Splitting problem.

Starting with our finite set $S$ and family of subsets $F $. 
$$ S =
  \left[ {\begin{array}{ccccc}
    a_{1} & a_{2} & a_{3} & a_{4} & a_{5}
  \end{array} } \right]
$$
$$
  F =
  \left[ {\begin{array}{ccccc}
    a_1 & a_2\\
    a_2 & a_4 \\
    a_1 & a_3& a_5\\
  \end{array} } \right]
$$
$$
X = \bordermatrix{ & a_{1} & a_{2} & a_{3} & a_{4} & a_{5} \cr
      S & ? & ? & ? & ? &? \cr} \qquad
$$
A clear solution is $S_1 = (a_2, a_3, a_5)$, $S_2 = (a_1, a_4)$ and it would be encoded in the matrix below.
$$
X = \bordermatrix{ & a_{1} & a_{2} & a_{3} & a_{4} & a_{5} \cr
      S & 0 & 1 & 1 & 0 & 1 \cr} \qquad
$$
$$\text{or}$$
$$
X = \bordermatrix{ & a_{1} & a_{2} & a_{3} & a_{4} & a_{5} \cr
      S & 1 & 0 & 0 & 1 & 0 \cr} \qquad
$$
This is the nature of the QUBO formulation, it has 2 unique yet valid solutions. This isn't a problem, as regardless of which solution is measured, the same information is conveyed.

Lets see how this could be encoded into a Quantum Annealer. We'll start by constructing an upper-triangle encoding matrix $Q$ to denote the biases between qubits.

\begin{equation} 
 H(X) = \sum_{j=0}^{n-1} \frac{1}{m-1}\sum_{i_0=0}^{m-1} \sum_{i_1=i_0+1}^{m-1} X_{F(j,i_0)} X_{F(j,i_1)} + \sum_{j=0}^{n-1} \frac{1}{m-1}\sum_{i_0=0}^{m-1} \sum_{i_1=i_0+1}^{m-1} (1-X_{F(j,i_0)})(1-X_{F(j,i_1)})
\end{equation}

For $n=3$, $m=[2,2,3]$,and our defined set $F$ we can simplify $H(X)$ to:
\begin{equation}\label{H5}
 H(X) =2x_1x_2+2x_2x_4+\frac{1}{2}(2x_1x_3+2x_1x_5+2x_3x_5) -2x_1-2x_2-x_3-x_4-x_5 
\end{equation}
 
 $$Q =
  \left[ {\begin{array}{cccccccccc}
    -2 &2& 1& 0& 1\\
     & -2& 0& 2& 0  \\
     & & -1& 0& 1  \\
     & && -1& 0\\
    & & & & -1\\
  \end{array} } \right]
 $$
\section{Results}
The above approach was tested with various $k=2$ set-split sizes on the D-Wave Advantage$-$system 4.1 Solver which takes advantage of the Pegasus topology. The QUBO module was encoded and run using the Ocean SDK. Table 1 displays the number of elements in a given problem as well as the number of Logical qubits (the minimum theoretically needed), Physical qubits (the amount used to encode the QUBO), and the total access time of the QPU which includes programming time and the sampling time for that particular job.


\begin{table}[h!]
\begin{center} 
\caption{Job information by Set Size} \label{PAtable} 
\vskip 0.005in
\begin{tabular}{lrrrrrr} 
\hline
\textbf{Problem Size}    & \textbf{Logical Qubits}   &   \textbf{Physical Qubits} & \textbf{QPU access time ($\mu s$)}\\
\hline
5 &   5 & 5 & 24499.57  \\
10 &    10 & 10 & 26084.77  \\
50 &    50 & 161 & 33425.17 \\
75&    75 & 341 & 29409.97   \\
100 &   100 & 513 & 33447.97  \\
200 &    200 & 2313 & 40148.37 \\
250 &    250 & 3348 & 41244.37   \\
300 &    300 & 4343 & 41759.97  \\
\hline
\end{tabular}  
\end{center} 
\end{table}
Here the tested problems were constructed by first generating a random target solution $X$ of set size - where the size of a solution is the number of elements in that solution - next, F was constructed of size $k=2$ subsets such that the generated solution was the only valid solution. This approach does lead to more Physical qubits being used than may be required for another problem of the same size but it ensures consistency among generated problems. A down side of physical D-Wave QPU Quantum Annealing hardware is that it is not fully connected, therefore because of how large F is, the number of qubits that require biases is very large, so the number of Physical qubits used in embedding scale much more quickly with problem size (seemingly exponential) than the theoretical qubits required.  However current restrictions mean that additional embedding qubits are required to couple two separate qubits. Unfortunately this both increases the number of qubits necessary for a given problem size, and increases the chain length, which decreases the stability of the result. This does imply that future embedding methods will be able to reduce the required qubits significantly.

\begin{figure}[ht]
    \centering
    \includegraphics[width=0.5\linewidth]{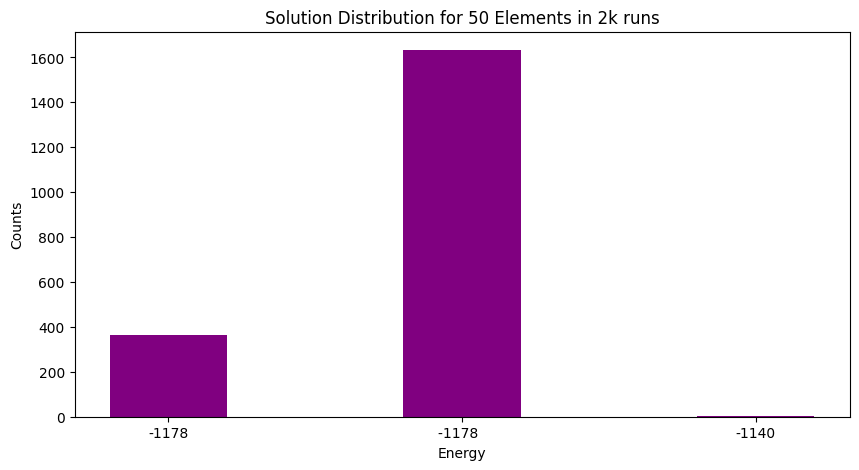}
    \caption{Energy state solutions (x-axis) versus occurrence in 2000 runs }
    \label{fig:enter-label}
\end{figure}
Despite stability concerns, the reliability of the constructed $k=2$ QUBO for $50$ elements is extremely high. We can see in Figure 1 that the minimum energy solution (Energy = $-1178$) is found in $\frac{1999}{2000}$ cases, and the false positive rate is consistently extremely low across all tested problem sizes. Given current qubit topology limitations, there is a practical threshold where larger problem sizes would lead to extremely high chain lengths and thus lower reliability of results. The max chain length with respect to problem size for both $k=2$ and $k=3$ can be seen in Figure 2.
\begin{figure}[ht]
    \centering
    \includegraphics[width=0.5\linewidth]{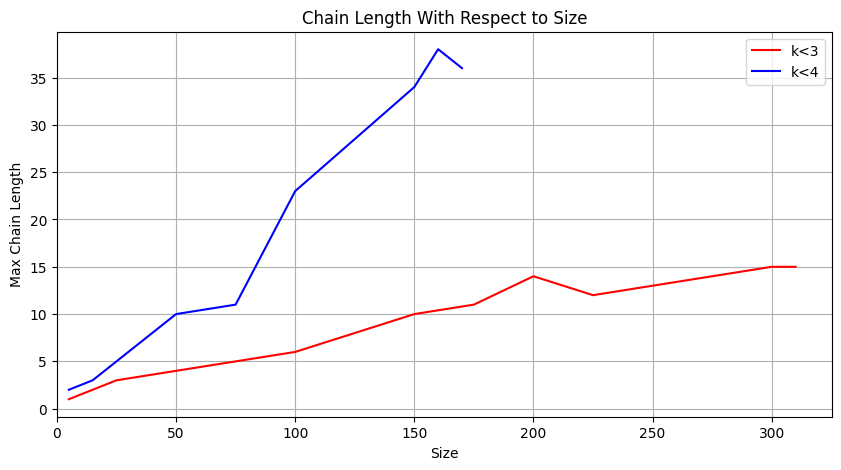}
    \caption{Chain Length over various problem sizes}
    \label{fig:enter-label}
\end{figure}

\section{Discussion}
This paper presents a novel method of solving or optimizing the Set Splitting Problem which utilizes quantum annealing to optimize a cost function encoded in a QUBO. The Logical bits scale linearly with the problem size and the solution is shown to have a low error rate for a limited test size. In terms of  time complexity the theoretical minimum, assuming no additional encoding qubits, is $O(e^{\sqrt{n}})$ since the number of qubits $n$ is equal to the number of elements. The proposed solution can solve the Set Splitting problem as well as k-Set-Splitting Problem and Weighted Set Splitting Problem with no issue for sets with a cardinality of $k \le 3$.  For $k \ge 3$ the solution is vulnerable to a condition where the energetic ground state may not correspond to a true solution; this is a trade off incurred for keeping with the QUBO model although it is a limitation.

There are a few key advancements that should be further researched. More expansive tests for larger $k$ values and for problems with more solutions would provide insight as to the limitations of this approach. I hypothesize that for both larger $k$ values and larger problem sizes the required physical qubits continues to increase exponentially. However this nonlinear growth is solely due to the current qubit connection limitations, and with Quantum Computers still in their infancy it's impossible to say if this exponential qubit requirement will last, or whether it might be subject to a type of Moore's Law improvement.

Generally, the set-splitting problem provides a flexible graph-based framework for formulating any combinatorial optimization challenge, where the interdependencies among the variables make exhaustive search of the solution space intractable. The Set-Splitting problem has usages in cybersecurity and network organizaition and can also be use to analyze micro array data and generate gene phenotype expressions. Formulating such problems as a set-splitting problem allows one to leverage the quantum system's parallelism for solving systems-level optimization problems, and for dynamically adjusting constraints and priorities as real-time conditions change.

\newpage
\addcontentsline{toc}{section}{References}
\printbibliography{}

\end{document}